\documentclass{sig-alternate}

\usepackage{url}

\newtheorem{theorem}{Theorem}
\newtheorem{defn}{Definition}

\begin{document}

\setlength{\pdfpageheight}{\paperheight}
\setlength{\pdfpagewidth}{\paperwidth}

\title{A likelihood based framework for assessing network evolution models tested on real network data}

%
%
%
%
%

\numberofauthors{4} 
%
\author{
\alignauthor Richard G. Clegg \\
       \affaddr{Department of Electrical and Electronic Engineering}\\
       \affaddr{University College London}\\
       \affaddr{London, UK}\\
       \email{richard@richardclegg.org}
\alignauthor Raul Landa\\
       \affaddr{Department of Electrical and Electronic Engineering}\\
       \affaddr{University College London}\\
       \affaddr{London, UK}\\
       \email{rlanda@ee.ucl.ac.uk}
\alignauthor Uli Harder\\
       \affaddr{Department of Computing}\\
       \affaddr{Imperial College London}\\
       \affaddr{London, UK}\\
       \email{uh@doc.ic.ac.uk}
\and
\alignauthor Miguel Rio\\
       \affaddr{Department of Electrical and Electronic Engineering}\\
       \affaddr{University College London}\\
       \affaddr{London, UK}\\
       \email{m.rio@ee.ucl.ac.uk}
}

\date{7 May 2007}

\conferenceinfo{SIMPLEX'09,} {July 1, 2009, Venice, Italy.} 
\CopyrightYear{2009}
\crdata{978-1-60558-704-2/09/07}

\maketitle
\begin{abstract}
This paper presents a statistically sound method for using likelihood to assess potential models of network evolution.  The method is tested on data from five real networks.  Data from the internet autonomous system network, from two photo sharing sites and from a co-authorship network are tested using this framework.
\end{abstract}




\section{Introduction}
\label{sec:Introduction}
It has been found that networks arising in very different contexts share
some structural statistical properties, for example a power law in their degree
distribution.
Such networks include the Internet Autonomous System (AS) topology, the WWW
hyperlink graph, co-authorship networks, sexual contact networks, social
networks based on email exchange, biological networks and others.  
For examples and references see \cite[table 3.1]{handbook}

One common hypothesis for the basis of these shared characteristics is the
presence of common elementary network development processes, such as the
{\em preferential attachment\/} model of Bar\'{a}basi et al. \cite{ba}. Other
models have been proposed for the evolution of specific classes of
networks. Many authors have proposed models which attempt to explain the
evolution of a target network in terms of simple rules which 
produce artificial networks with the same characteristics as a given target
network. Examples of models of
this kind can be found in 
\cite{ba2,ba,bu,zhou2004}.  In the literature, such models
are usually tested by growing an artificial model of the same size
as the target network and comparing several network statistics on
the real and artificial networks.

In this paper we propose the Framework for Evolutionary Topology Analysis
(FETA).  This framework provides several advantages when compared
to the usual method of testing models: a single likelihood
based measure of how well a proposed model explains the observed
network evolution, it uses network evolution data rather
than a single static network snapshot, and it includes
a method for creating new network models from linear
combinations of sub-models and a method for optimising the mixture 
of these sub-models.  The FETA allows the assessment of network models
without growing artificial models and comparing them to the target network, making 
model testing much faster.
This paper is a companion to \cite{clegg09} which introduced the framework
and showed it could recover known
model parameters for artificial network models.  
This paper shows that the FETA framework
can be used to investigate a variety of real networks.
The class of models which FETA can work with includes Barab\'{a}si--Albert (BA)
\cite{ba}, Albert--Barab\'{a}si (AB) \cite{ba2}, Generalised Linear Preference (GLP) 
\cite{bu} and Positive Feedback Preference (PFP) \cite{zhou2004}.

\section{A likelihood based framework for assessing network models}
\label{sec:method}
The probabilistic models used by FETA
are described in terms of two components referred to as
an {\em inner model\/} and an {\em outer model\/}.  

\begin{defn}
The {\em outer model\/} chooses the operation that
transforms the current 
network.  This could be add a node, add a link
between existing nodes, delete a node or delete a link between nodes.
\end{defn}

\begin{defn}
The {\em inner model\/} chooses the entity on which the operation will
act.  More simply it defines a probabilistic model which gives the
probabilities of choosing nodes or links for the add or remove operation
selected by the outer model.
\end{defn}

A simple example would be the AB model.  This would correspond
to an outer model which adds 
a new node and then chooses exactly three inner nodes to connect to it.  The
inner model assigns probabilities to each inner node exactly proportional 
to  their node degree. 
As is common in the literature, the main focus of FETA is on
the inner model.  
The framework is flexible enough to allow or disallow node and edge
removal, non-simple and directed graphs.  For this paper, however, only
connected, simple, undirected graphs which never lose nodes
or edges are considered.

Let $G_0$ 
be the known state of the graph at a certain time.
Assume that graph is known for each time
an edge is added up to some step $t$ ($G_0, G_1, \ldots, G_t$ is 
known).  Let $\theta$ be a proposed model which attempts to
explain this evolution.  The model $\theta$ assigns probabilities
to entities in the network at each step of the network evolution.

In order to simplify the explanation, 
assume that the outer model always
involves the choice of a single existing node to connect to a new
node.  Let $C = (N_1,\ldots,N_t)$ be the ordered list of nodes selected
at each step derived from $G_0, \ldots, G_t$.
Let $p_i(j | \theta)$ be the probability that inner model
$\theta$ assigns to node $j$ at step $i$ -- that is, the probability that node
$j$ is chosen at step $i$.  To be a valid model
$\theta$ should ensure that $\sum_j p_i(j | \theta) = 1$ where
the sum is over nodes.  The following theorem can easily be shown 
\cite{clegg09}.

\begin{theorem}\label{thm}
Let $C=(N_1,\ldots,N_t)$ be the observed node choices at steps $1,\ldots,t$ of
the evolution of the graph $G$.  
Let $\theta$ be some hypothesised valid inner model which assigns
a probability $p_j(i|\theta)$ to node $i$ at step $j$.  The likelihood of
the observed $C$ given $\theta$ is
$$
L(C|\theta) = \prod_{j=1}^t p_j(N_j|\theta).
$$
\end{theorem}

Note that the probability $p_j(i|\theta)$ may depend on many things including
past history, node properties exogenous to the graph and previous node
choices.  As long as these are observable then the calculation is still easy
to make.  The next step is to define a null model $\theta_0$ to compare
the hypothesised model $\theta$ against.

\begin{defn} \label{defn:c_0}
The {\em null model\/} $\theta_0$ is defined as 
the model which gives every
node in the choice set equal probability (this can also be
thought of as the {\em random model\/}). 
The {\em per choice likelihood ratio\/} $c_0$ is the likelihood ratio between
$\theta$ and the null model $\theta_0$ normalised by the number of choices.
$$c_0= \left[\frac{L(C|\theta)}{L(C|\theta_0)}\right]^{1/t}.$$
\end{defn}

The quantity $c_0$ is one if $\theta$ is exactly as likely as $\theta_0$ to have
given rise to the observed choices $C$.  If $c_0$ is greater
than one then $\theta$ is more likely and if less than one it is less likely.
Note that two hypothesised models can be compared by looking at the ratio of their
$c_0$ values.
Note also that $c_0$ and $L(C|\theta)$  are simply different ways of
looking at the model likelihood.  
It is also worth noting that, using the same underlying
source code to calculate the probabilities, 
generating an artificial network model of the same size
as the real target network took much longer (sometimes
a hundred times as long) than measuring the likelihood statistics.
Other standard statisitics such as deviance and Akakai's Information
Criterion can also trivially be calcualted from $L(C|\theta)$.

If $\theta_i$ $i=1,2,\ldots,N$ are valid models for a given network
then $\theta=\sum_{i=1}^N \beta_i \theta_i$ with $\beta_i \in [0,1]$ and
$\sum_{i=1}^N \beta_i =1 $ is also a valid model.  This allows sub models
to be linearly combined to form hybrid models.  The linear $\beta$ parameters and
other model parameters (for example the $\delta$ in the PFP model) can be
optimised to find the model which has the highest $c_0$ value for a given
target network.  Optimisation of the $\beta$ parameters can be performed
using generalised linear modelling as described in \cite{clegg09}.

Let $d_i$ be the degree of node $i$ and $T_i$ be the triangle
count (the number of triangles, or 3--cycles, the node
is in).
The model components considered for this paper included the following:
$\theta_0$ -- the null model (random model) assumes all nodes have equal 
probability $p_i = k_0$;
$\theta_d$ -- the degree model (preferential attachment) 
assumes node probability $p_i = k_d d_i$;
$\theta_T$ -- the triangle model assumes node 
probability $p_i = k_t T_i$;
$\theta_S$ -- the singleton model assumes node 
probability $p_i = k_S$ if $d_i = 1$ and $p_i = 0$ otherwise;
$\theta_D$ -- the doubleton model assumes node 
probability $p_i = k_D$ if $d_i = 2$ and $p_i = 0$ otherwise;
$\theta_R(n)$ -- the ``recent" model where $p_i = k_H$ if a node
was one selected in the last $n$ selections and $p_i = 0$ otherwise
and
$\theta_p^{(\delta)}$ -- the 
PFP model assumes node 
probability $p_i = k_p d_i^{1 + \delta \log_{10}(d_i)}$.
The $k_{\bullet}$ 
are all normalising constants to ensure $\sum_i p_i = 1$.

So, for example $\theta= 0.5 \theta_d + 0.4 \theta_p(0.05) + 0.1 \theta_S$
is a model which is 50\% preferential attachment, 40\% PFP with $\delta=0.05$
and 10\% singleton model.

\section{Real data testing}
\label{sec:real}
The FETA procedure is used to create inner models for several networks of
interest.  Section \ref{sec:pub} fits models to a co-authorship network 
inferred from the arXiv database.  Section \ref{sec:ucla} fits models to a
view of the AS network topology referred to here as the UCLA AS network
and section \ref{sec:routeviews} fits models
to a second view of the AS topology, which we refer to here as the RouteViews
AS network.
Section \ref{sec:gallery} fits network evolution models 
to a network derived from user browsing behaviour on a photo
sharing site known as ``gallery" and 
section \ref{sec:flickr} fits models to a social network derived 
from the popular photo sharing site Flickr. 
The networks are summarised below.
\begin{center}
\begin{tabular}{|c|l l l|} \hline 
Network & edges & nodes & edge/node  \\ \hline
arXiv & 15,788 & 9,121 & 1.73 \\
UCLA AS & 93,957 & 29,032 & 3.24\\
RouteViews AS & 94,993 & 33, 804& 2.81 \\
gallery &  50,472  & 26,958  & 1.87 \\
Flickr & 98,931 & 46,557 & 2.13\\ \hline   
\end{tabular}
\end{center}

For each data set, three inner models are tried: a random model,
a pure PFP model (with an optimally tuned $\delta$ for connections 
to new nodes, and another one for internal edges) and
the best model found by trying all combinations of submodels using
the generalised linear model fitting procedure described in
\cite{clegg09} and maximising the per choice likelihood ratio $c_0$ --
separate inner models are fitted to connections from new nodes and connections
between existing nodes.  These models will be called, for convenience,
random, PFP and {\em best\/} -- where best here should be understood as the best
possible model using combinations of the submodels consdered rather than being
the best possible model of the network.
Note that this model does not contain the interactive growth model
from \cite{zhou2004} and the results that follow should not be taken
as a criticims of PFP as described in \cite{zhou2004}.

Because the outer model was not the subject of interest here the
outer model was simply taken to be the actual operation observed
in the real data.  In practice this was little different from
the results obtained from the 
outer model derived simply by calculating empirically from the
data two distributions: 1) 
the number of inner nodes each new node connects to on arrival,
2) the number of inner edges connected between each new node arrival.
The outer model behaviour can be drawn from these distributions and
the results are little changed.

For each model, $c_0$ from definition \ref{defn:c_0} is measured.  Several
network statistics are then measured for comparison.  Simple statistics
were chosen: $d_1$ is the proportion of nodes which have degree one and
$d_2$ the proportion of nodes with degree two, $\max d$ is the maximum
degree of any node
and $\overline{d^2}$ is the mean square of the node degrees (a measure
of variance) --  note that $\overline{d}$ is not a useful measure, it is set by
the outer model and would be the same for all models.  
The clustering
coefficient  $\gamma$ 
is a measure of the proportion of possible triangles present in the graph.
The assortativity coefficient $r$ is positive when
nodes attach to nodes of like degree (high degree nodes
attach to each other) and negative when high degree
nodes tend to attach to low degree nodes.
For full definitions of all these quantities see 
\cite{hamedsurvey}.

\subsection{Fitting the arXiv data set}
\label{sec:pub}

A publication co-authorship network was obtained from the online academic
publication network arXiv\footnote{\url{http://www.arxiv.org}}.  
The first paper was added in April 1989 and
papers are still being added to this day.
To keep the size manageable, the network was produced just from the  papers 
with the category label ``math".  
The network is a co-authorship network: 
an edge is added when two authors first write a paper together. 
The author 
match is on first initial and surname, though it is
clear this will allow some collisions.
One paper\footnote{\url{http://arxiv.org/abs/math/0406190}} was removed from analysis.
The paper has 60 authors (far more than the next largest)
which would add a distorting size 60 clique (1,732 links).
The arXiv network has also been analysed by (amongst others)
\cite{leskovec2005} from the perspective
of growth rates and clique addition.

Obviously the random model has $c_0 = 1$.  The pure PFP model has 
$\delta= -0.17$ and $c_0= 1.31$.   The {\em best\/} model has the model
for connecting to new nodes  
$0.56\theta_p(-0.29) + 0.28\theta_R(3)
+0.16 \theta_S$ (PFP + recent + singleton)
and the model for connecting between existing nodes
$0.57\theta_p(-0.03) + 0.39\theta_R(3)
+0.04 \theta_S$ (PFP + recent + singleton) together this gives
$c_0 = 6.24$.  This implies that PFP should be slightly better
than random and {\em best\/} should be better than both.

\begin{figure*}[ht!]
\begin{center}
\includegraphics[width=5.5cm]{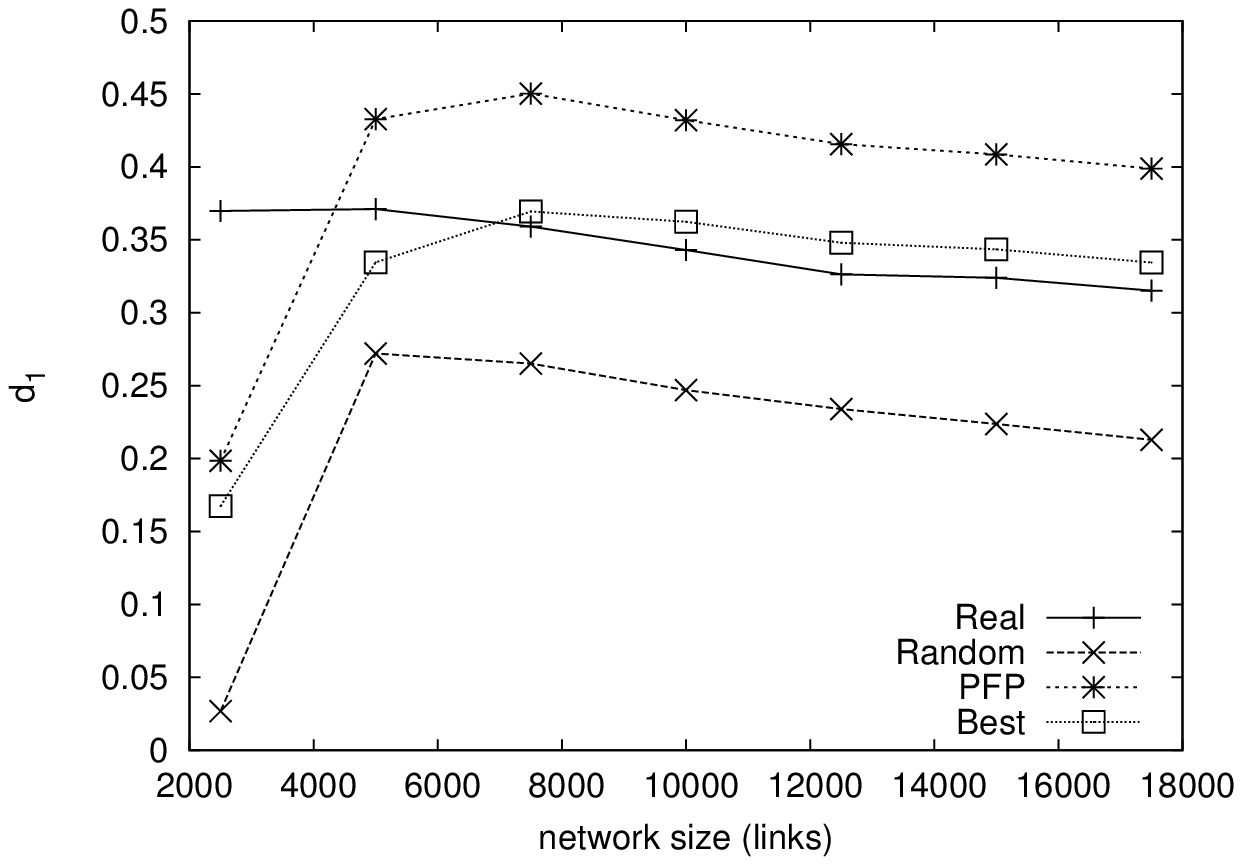}
\includegraphics[width=5.5cm]{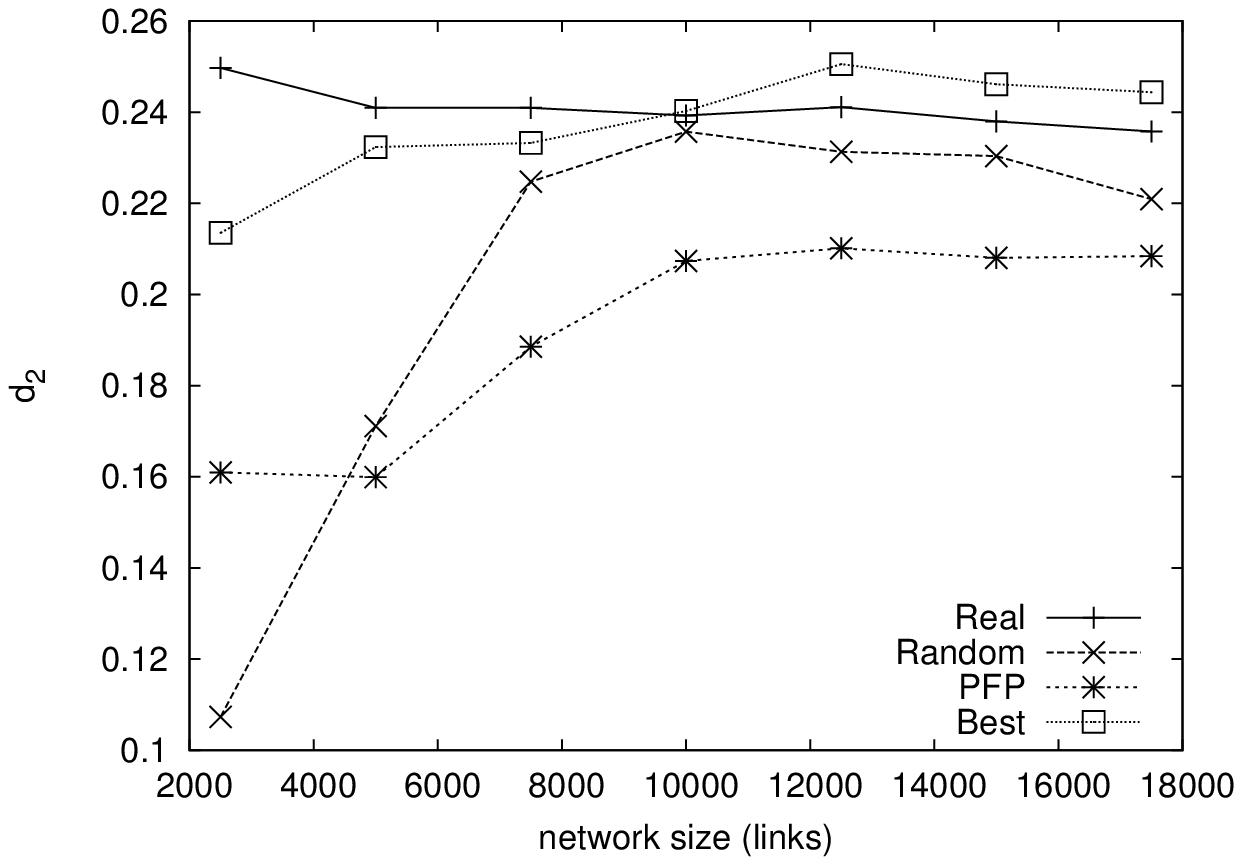}
\includegraphics[width=5.5cm]{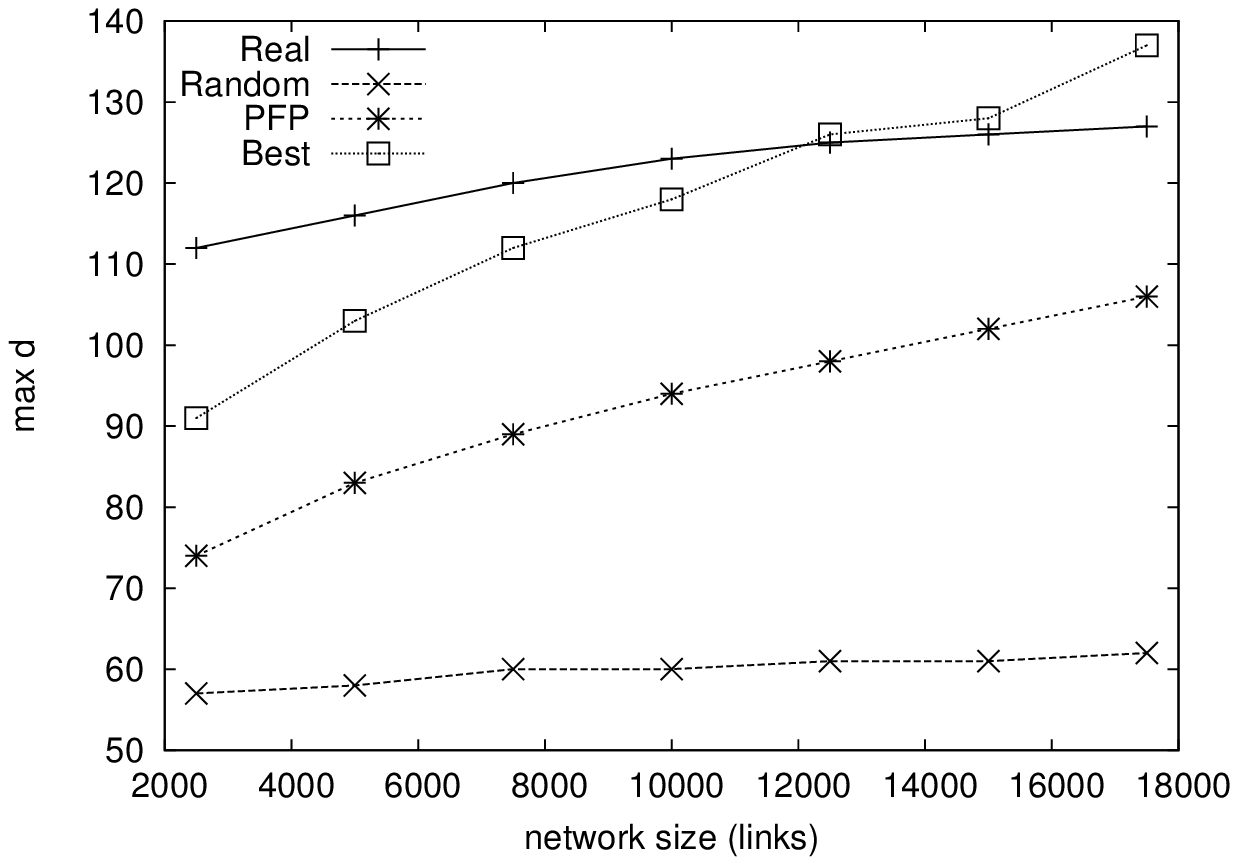} \\
\includegraphics[width=5.5cm]{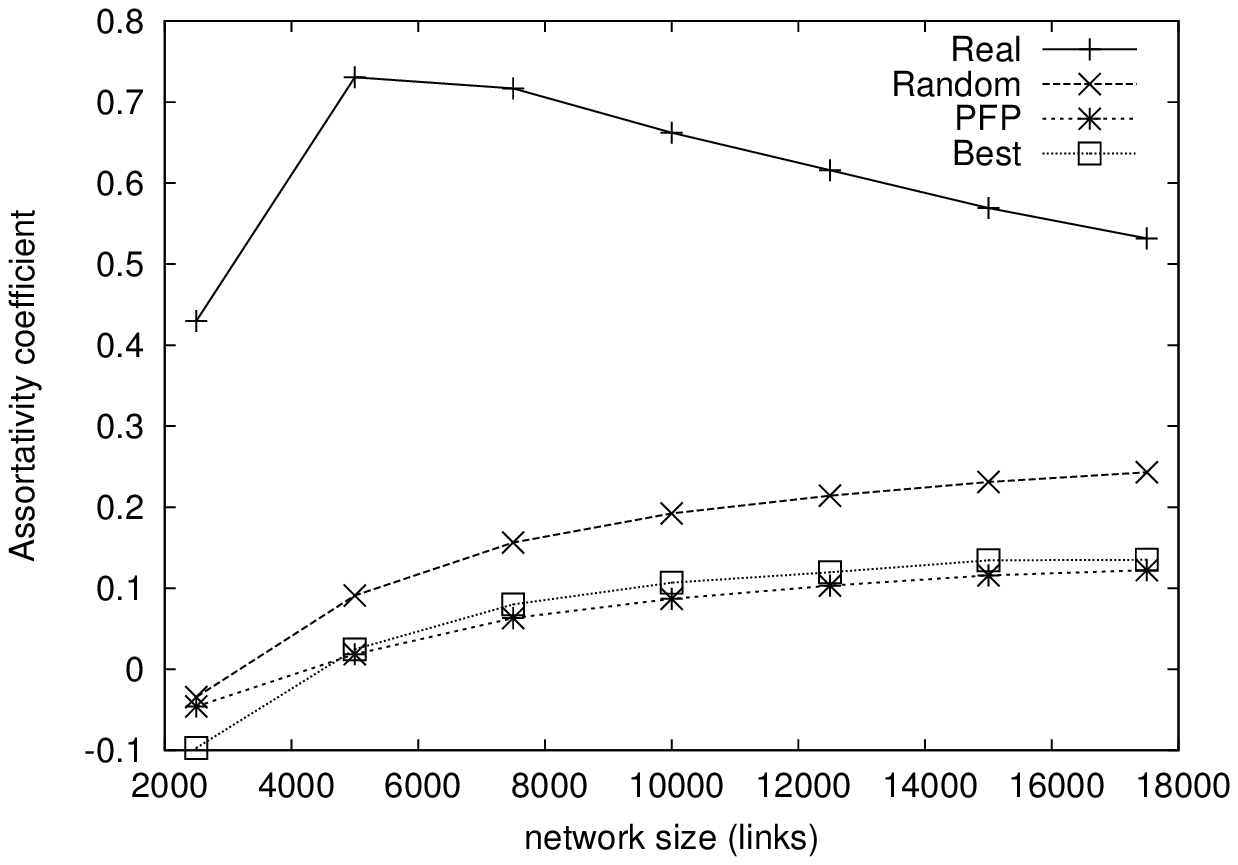}
\includegraphics[width=5.5cm]{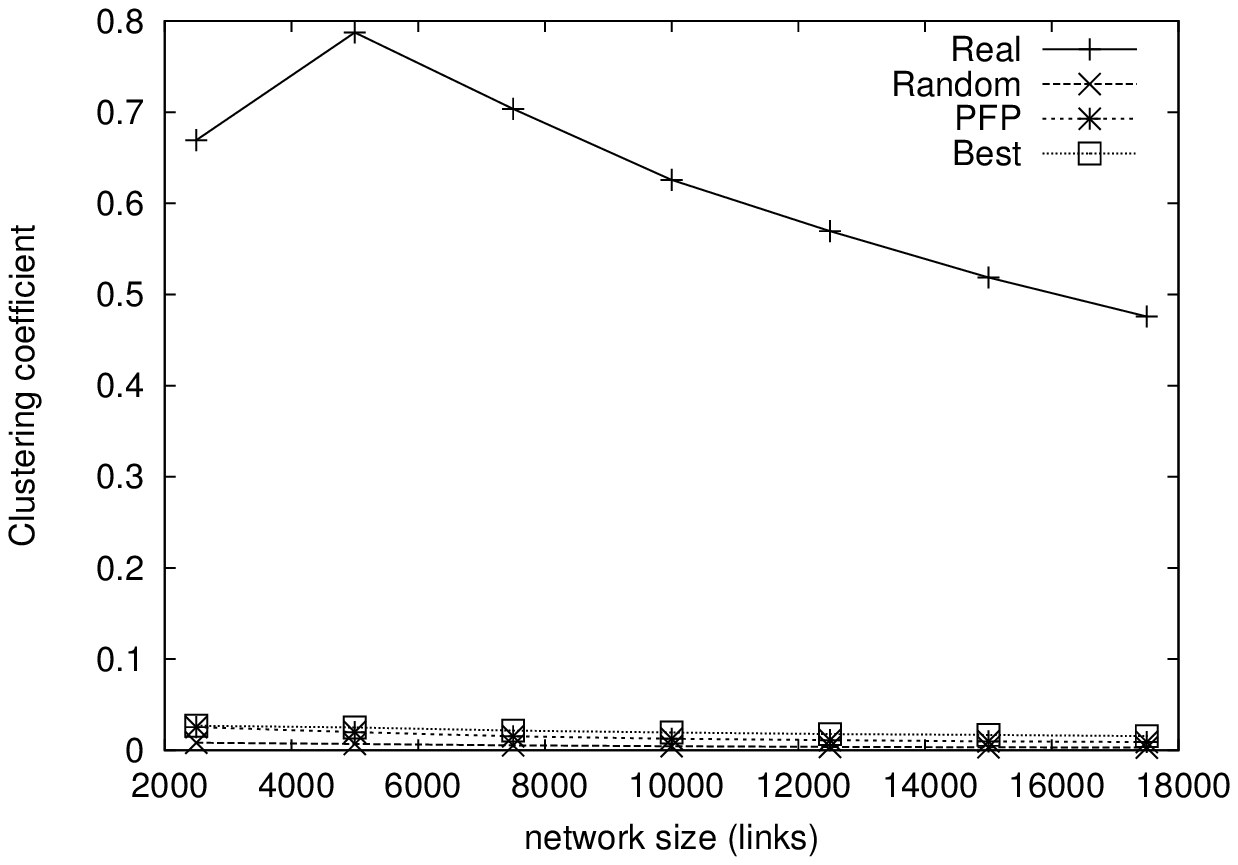}
\includegraphics[width=5.5cm]{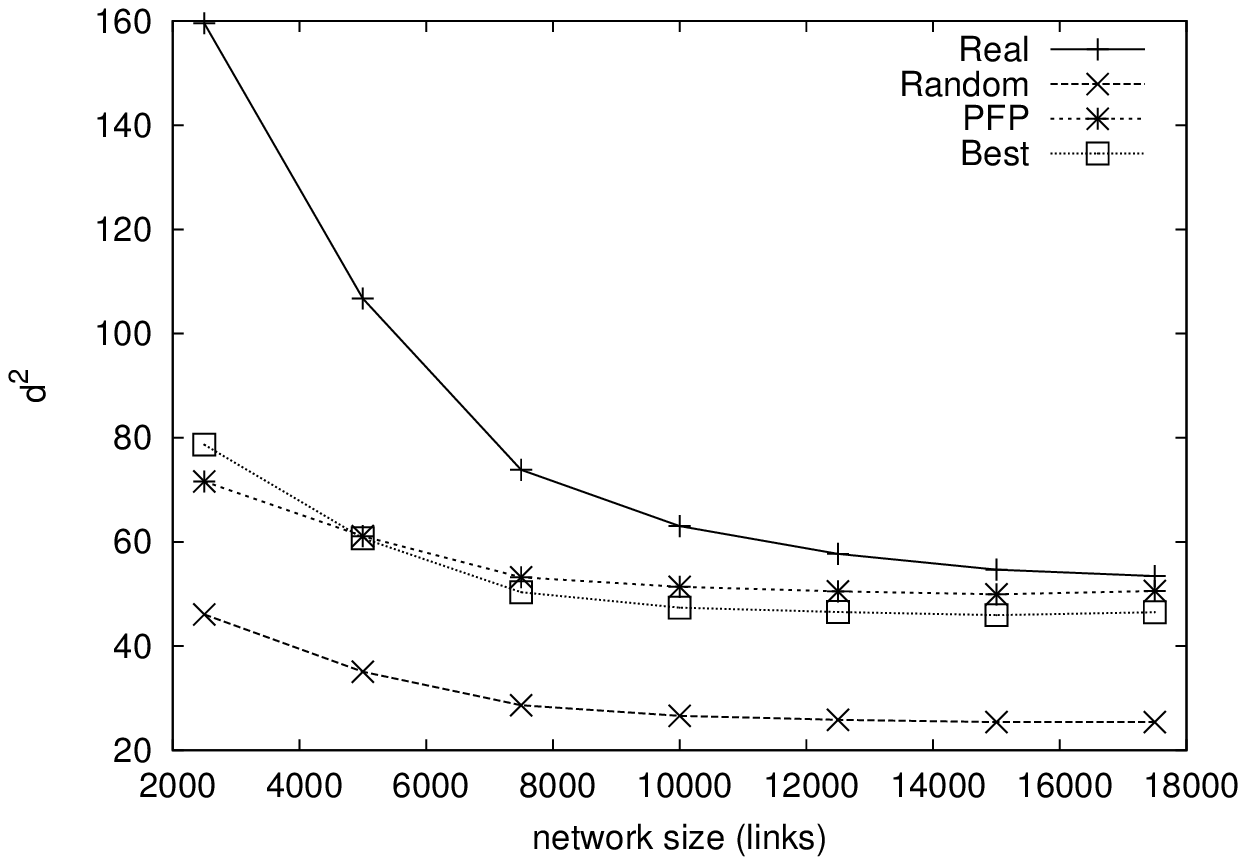}
\caption{Results for arXiv network}
\label{fig:arx}
\end{center}
\end{figure*}

Figure \ref{fig:arx} shows the results for the arXiv data.  As can be seen,
for $d_1$ and $\max d$ the results are in the order predicted
and, for the {\em best\/} model are a good fit to the real data.  For $d_2$ random
is slightly better than PFP.  For $\overline{d^2}$
PFP is a better fit than the {\em best\/} model although both are very similar and
quite close to the real data.  For $\gamma$ and $r$ all models are similar
and similarly bad fit to the real data.  No models have captured these
second and third order statistics.  The obvious reason for this is that
uniquely in the arXiv data nodes are all added as cliques.  If $n$ authors
write a paper together then a clique of size $n$ (some nodes in which
are already present on the network) is added.  An obvious improvement
to the model could be obtained by having ``add clique of size $n$" as an outer
model operation and an inner model which selected which node(s) in the clique
were already present in the network.

\subsection{UCLA AS data set}
\label{sec:ucla}

The data set we refer to here as the UCLA AS data set is a view of
the Internet AS topology 
seen between January 2004 and August 2008.  It comes from the Internet
topology collection\footnote{\url{http://irl.cs.ucla.edu/topology/}}
maintained by Oliviera et al. \cite{asevolution}. 
These
topologies are updated daily using data sources such as BGP routing
tables and updates from RouteViews,
RIPE,\footnote{\url{http://www.ripe.net/db/irr.html/}}
Abilene\footnote{\url{http://abilene.internet2.edu/}} and LookingGlass
servers. Each node and link is annotated with the times it was first
and last observed during the measurement period.  The AS data set has
been analysed by several other researchers but few have analysed
the data set as it grows.  \cite{dhamdhere2008} uses
linear modelling techniques to assess the goodness of fit of a preferential
attachment model.

The data is preprocessed by removing all edges and nodes
which are not seen in the final sixty days of the data, so that the
final state of the evolution of the network is the AS network as it is in August 2008.
Edges are introduced into the network in the order of their first sighting.
If this would cause the network to become disconnected, their introduction
is delayed.
Data is available from January 2004 and a ``warm up'' period is
given with
$G_0$ (the starting graph) taken to occur slightly after this start date.s

For the UCLA data the best pure PFP model was with 
$\delta = 0.0015$ which had $c_0 = 6.326$.  The {\em best\/} model
was, for the model to connect to new nodes 
$0.81\theta_p(0.0015) + 0.19\theta_R(1)$ and for the model
to connect between existing edges 
$0.75 \theta_d + 0.2 \theta_R(1) + 0.05\theta_S$.
This model had $c_0 = 11.43$.

\begin{figure*}[ht!]
\begin{center}
\includegraphics[width=5.5cm]{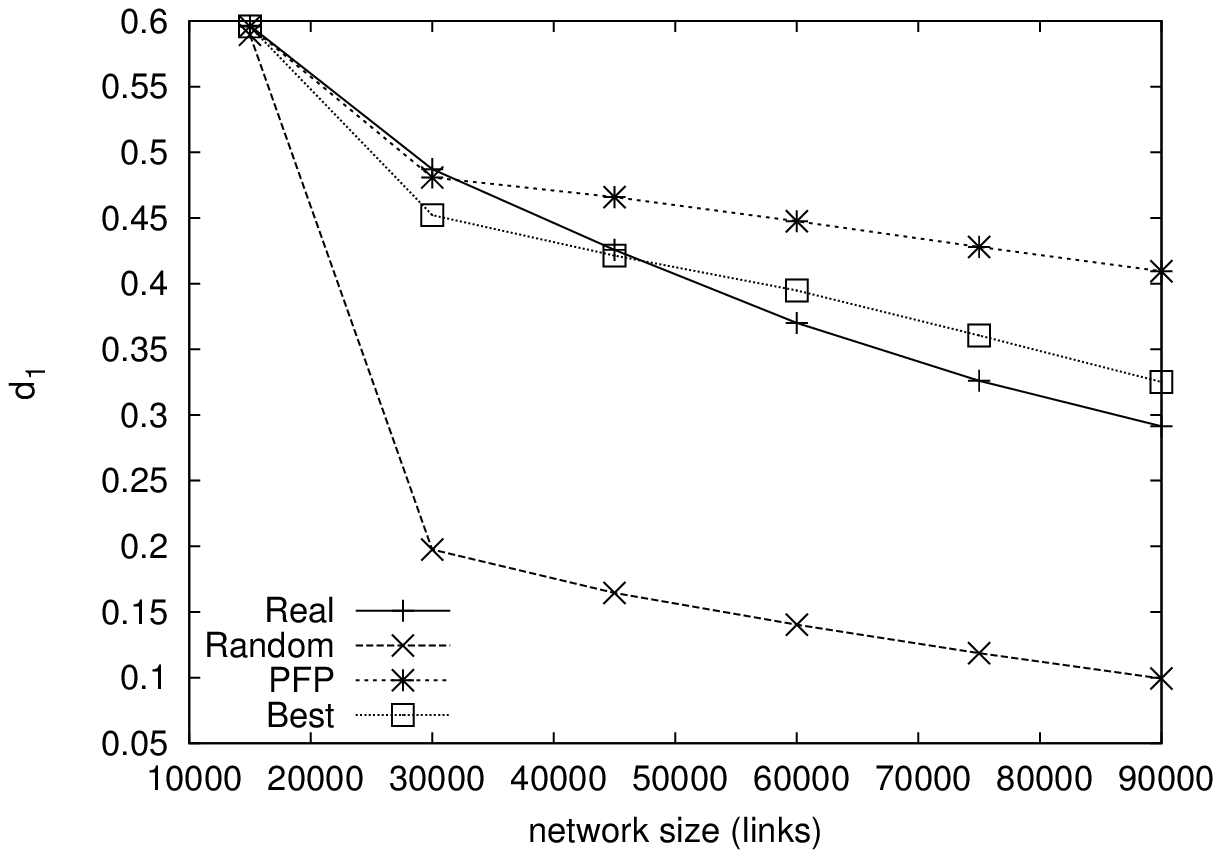}
\includegraphics[width=5.5cm]{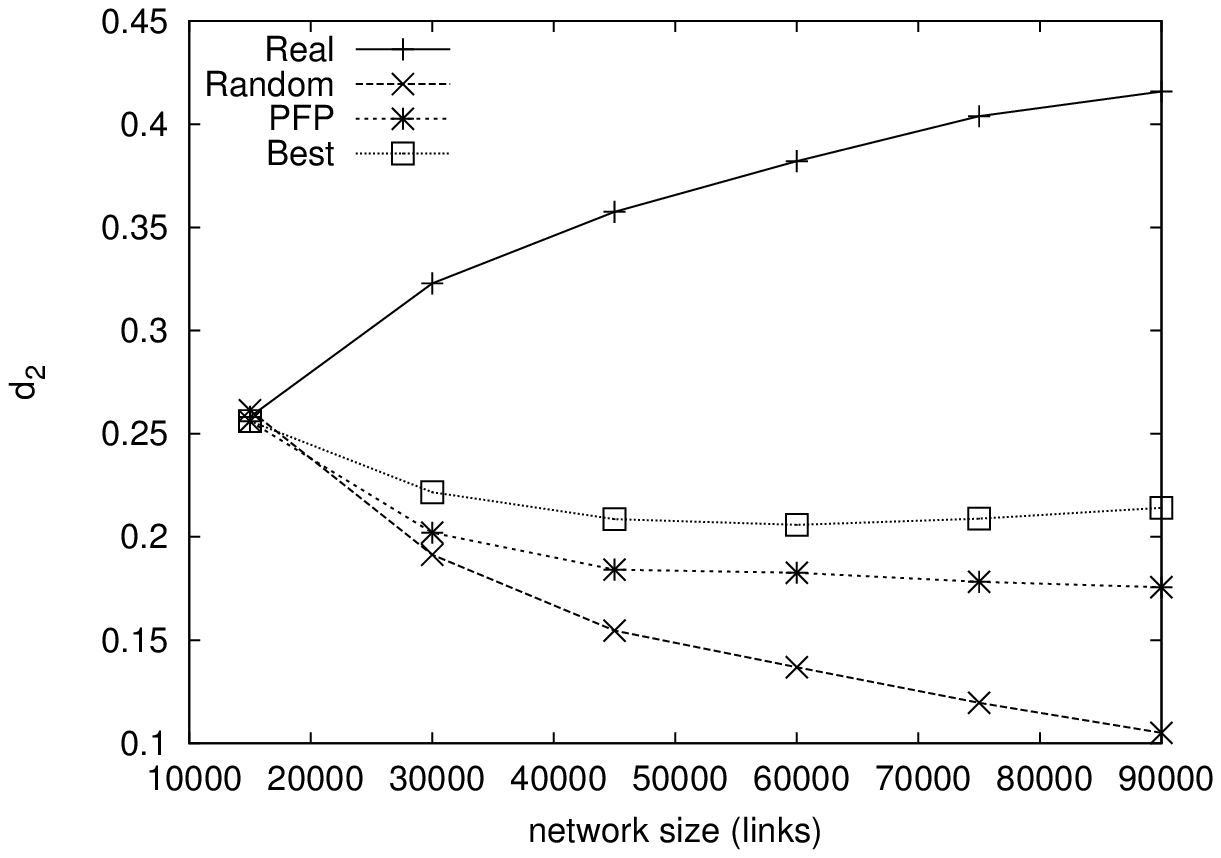}
\includegraphics[width=5.5cm]{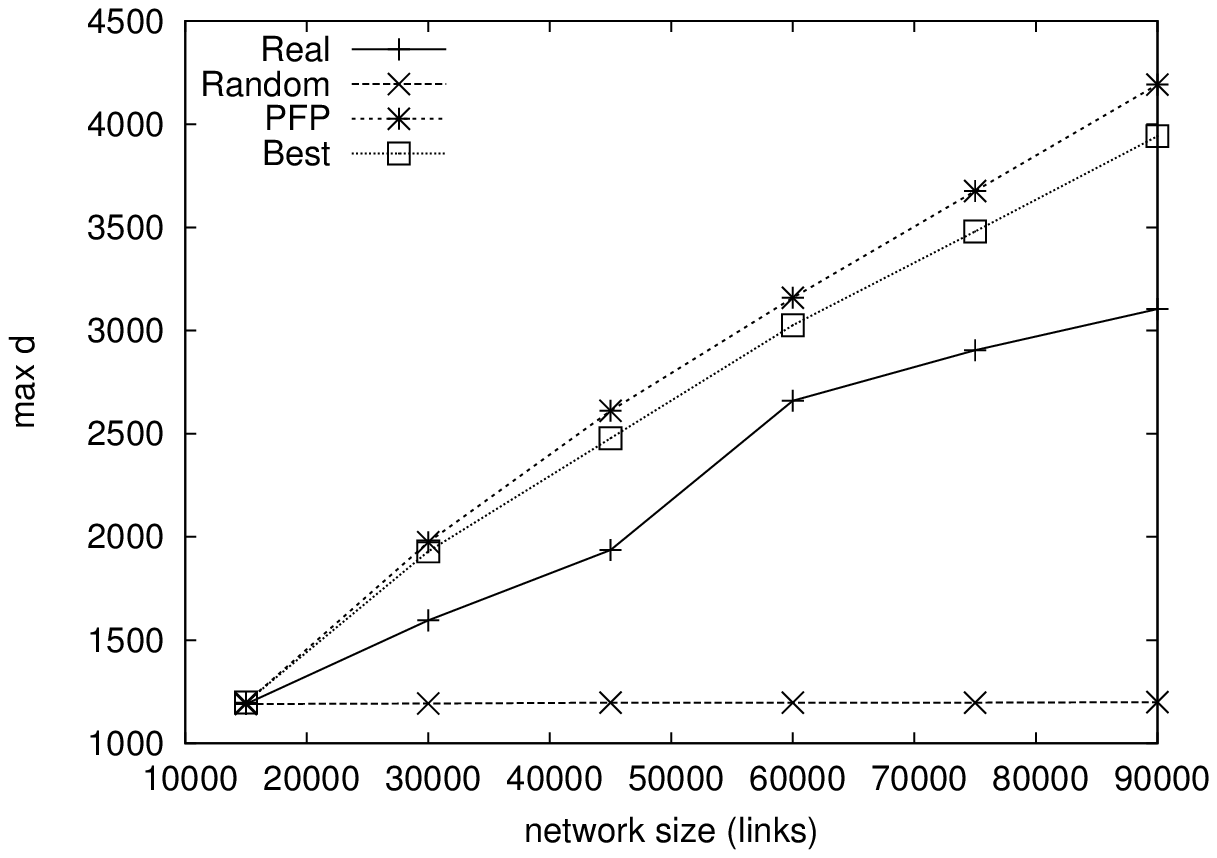} \\
\includegraphics[width=5.5cm]{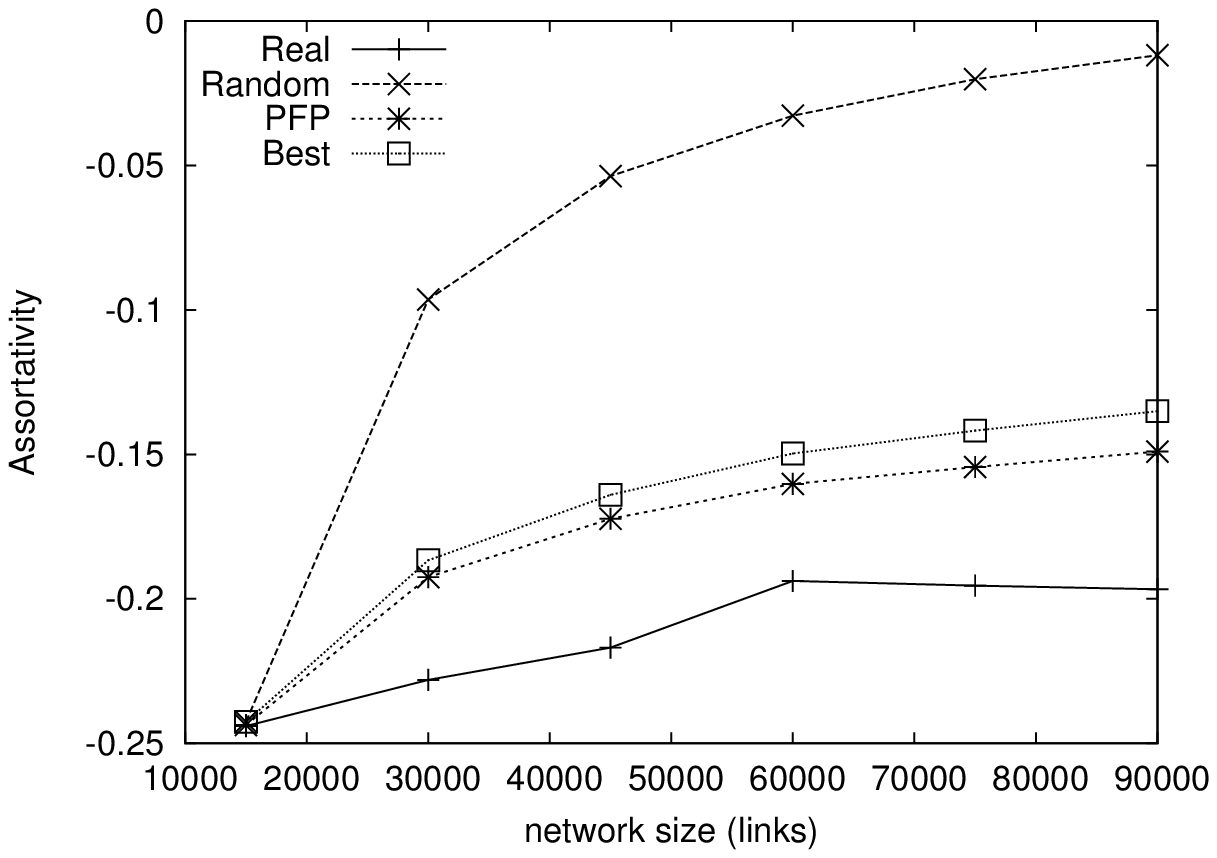}
\includegraphics[width=5.5cm]{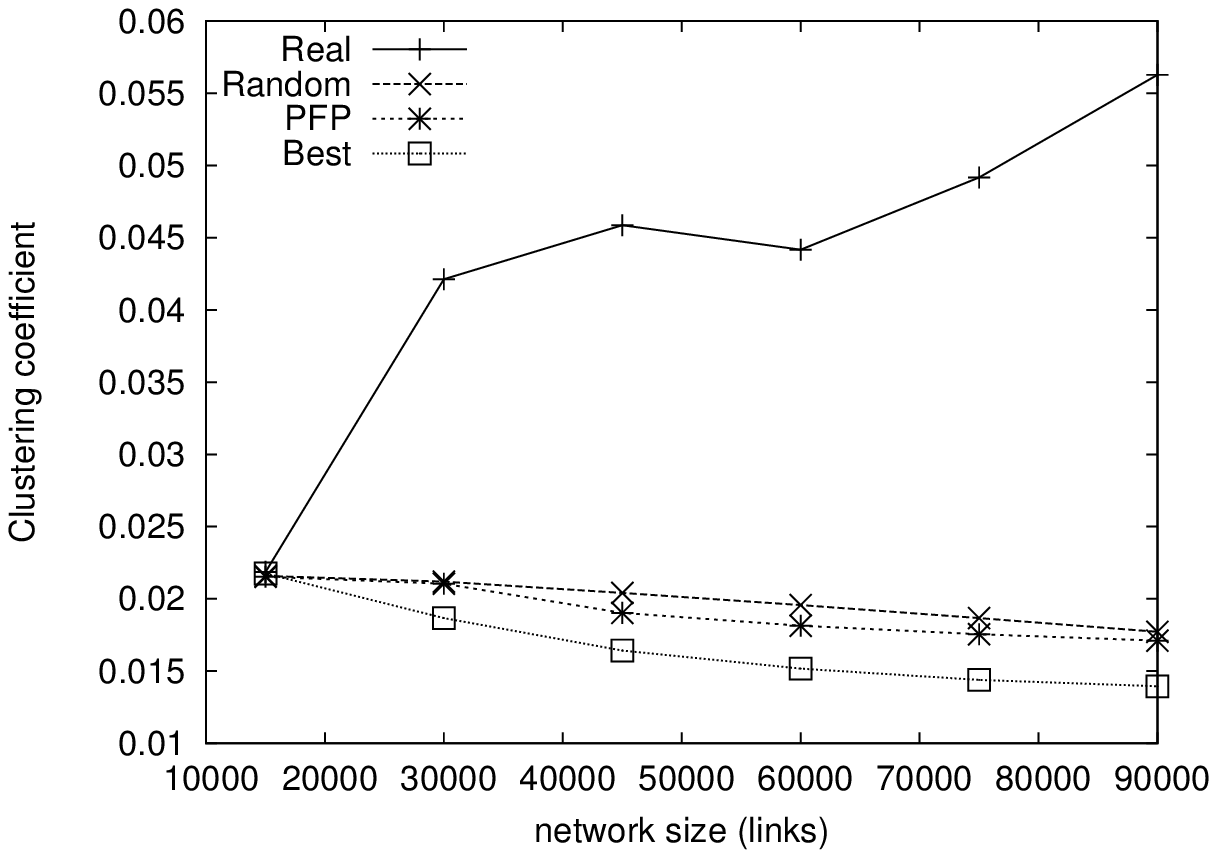}
\includegraphics[width=5.5cm]{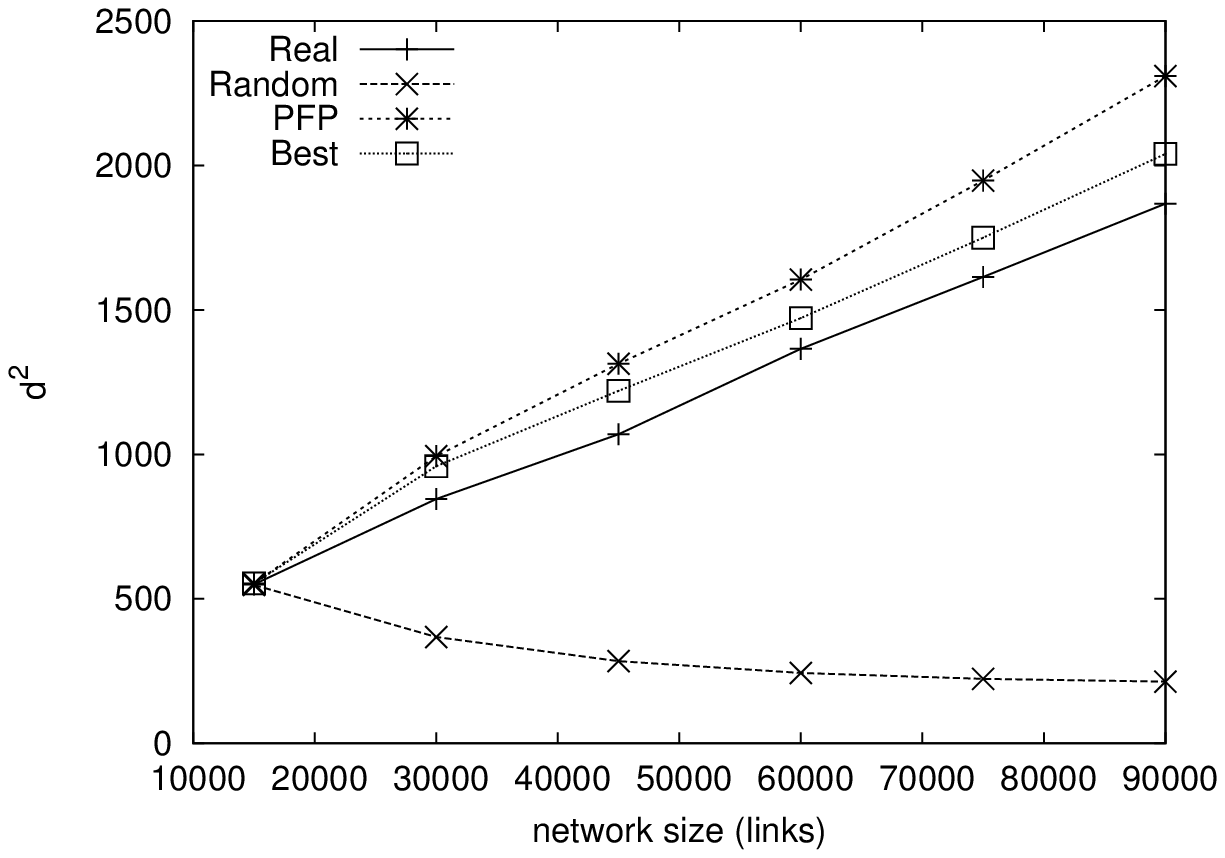}
\caption{Results for UCLA network}
\label{fig:ucla}
\end{center}
\end{figure*}

Figure \ref{fig:ucla} shows the results for the UCLA network.
For $d_1$, $d_2$, $\max d$ and $\overline{d^2}$ the results
are in the expected order and for all but $d_2$ are quite close
(no model predicts $d_2$ very well).  For assortativity, 
PFP is slightly better than {\em best\/}.  For clustering coefficient
no models are correct.

\subsection{RouteViews AS data set}
\label{sec:routeviews}

For the present paper we define the RouteViews AS data set 
as the view of the Internet AS topology 
from the point of view of a single RouteViews data collector.  The raw data 
used to construct it comes from the University of Oregon Route Views 
Project\footnote{\url{http://www.routeviews.org}}.  A fuller description
can be found in \cite{clegg09}. 
The best pure PFP model was $\theta_p(0.005)$ and the {\em best\/} 
model found which was $0.81\theta_p(0.014) + 0.17\theta_R(1)$ (PFP + ``recent") 
to connect new nodes
and $0.71\theta_d + 0.22\theta_R(1) + 0.07\theta_S$ (preferential attachment
+ ``recent" + singleton) to connect edges between
existing nodes.  
The PFP model $\theta_p(0.005)$
had $c_0 = 4.81$ and the
{\em best\/} model had $c_0 = 8.06$.  This suggested that {\em best\/} would be
better than PFP which would be better than
random.

For most statistics, the models were in the order expected but
for $\gamma$ and $r$
PFP was slightly better than {\em best\/}.  For $d_1$ and
$\max d$ the PFP model was little different to random
although in the case of $\max d$, random predicted unrealistically 
slow growth.   Overall, however, the model order was that predicted by
the $c_0$ values.

\subsection{Fitting the gallery data set}
\label{sec:gallery}

The website known simply as ``gallery"\footnote{\url{http://gallery.future-i.com/}}
is a photo sharing website. 
To be able to upload pictures and have some control over 
the display of pictures, users have to create an account and login.  
From webserver logs, the path logged in users browse as they move
across the network can be followed. Thus, images become nodes 
in the networks, and a user browsing between two
photos creates a link between the two nodes that represent them. 
These links are overlaid for all users in order to form the network
analysed here.

The best pure PFP model for the gallery data was with
$\delta = -0.4$, however, unusually, this model was worse than
random with $c_0 = 0.8515$.  The {\em best\/} model had, for its connections
to new nodes, $0.57\theta_S+0.24\theta_d+0.19\theta_R(3)$ (singleton +
preferential attachment plus ``recent") and for its connections
between existing nodes  $0.61 \theta_p(-0.05) + 0.39\theta_R(5)$.  This
model had a per choice likelihood ratio $c_0 = 12.93$.

\begin{figure*}[ht!]
\begin{center}
\includegraphics[width=5.5cm]{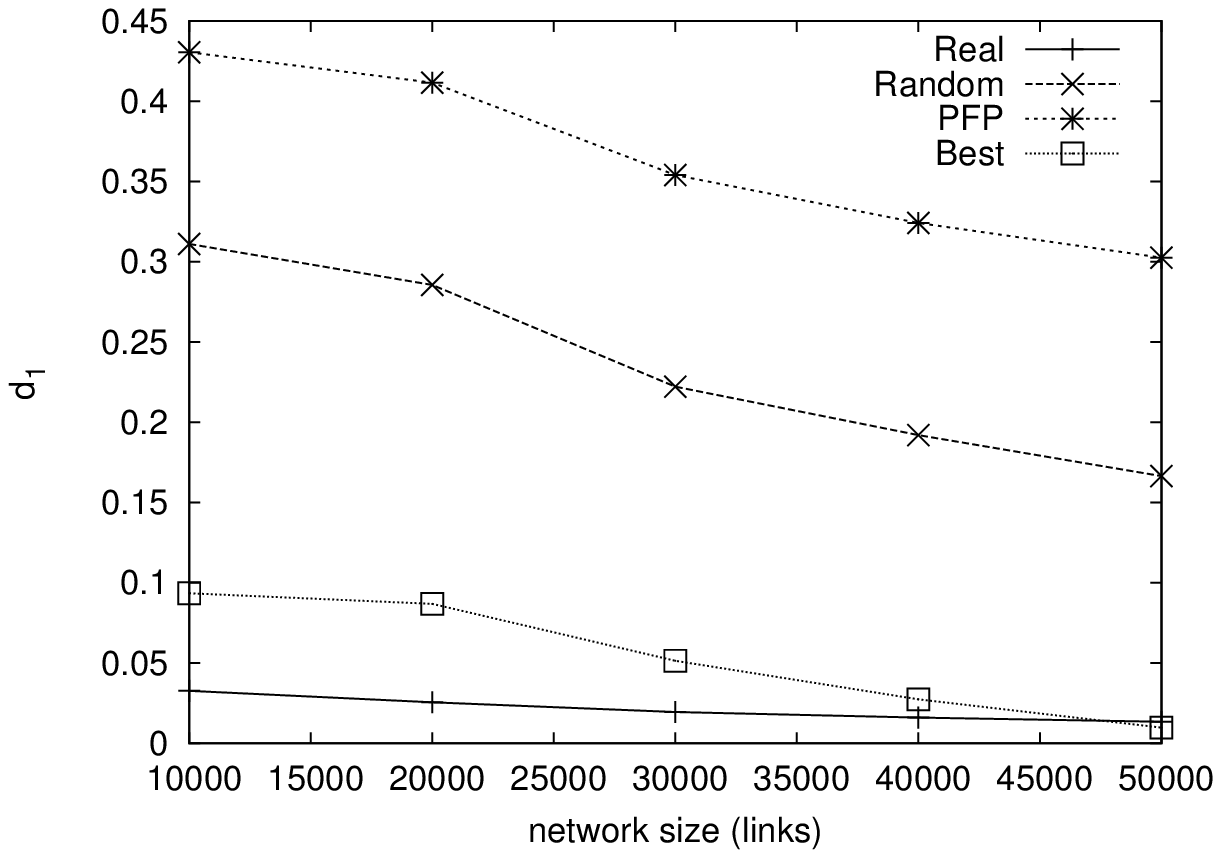}
\includegraphics[width=5.5cm]{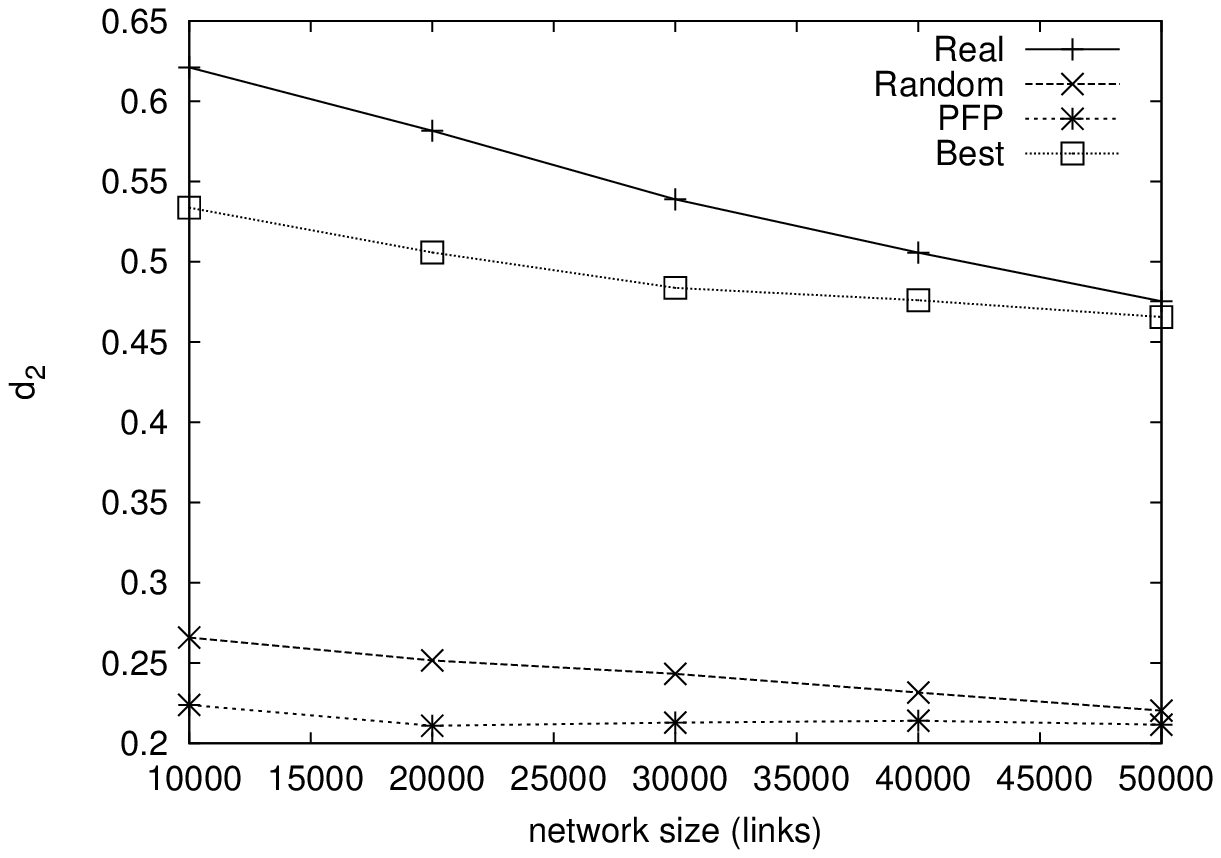}
\includegraphics[width=5.5cm]{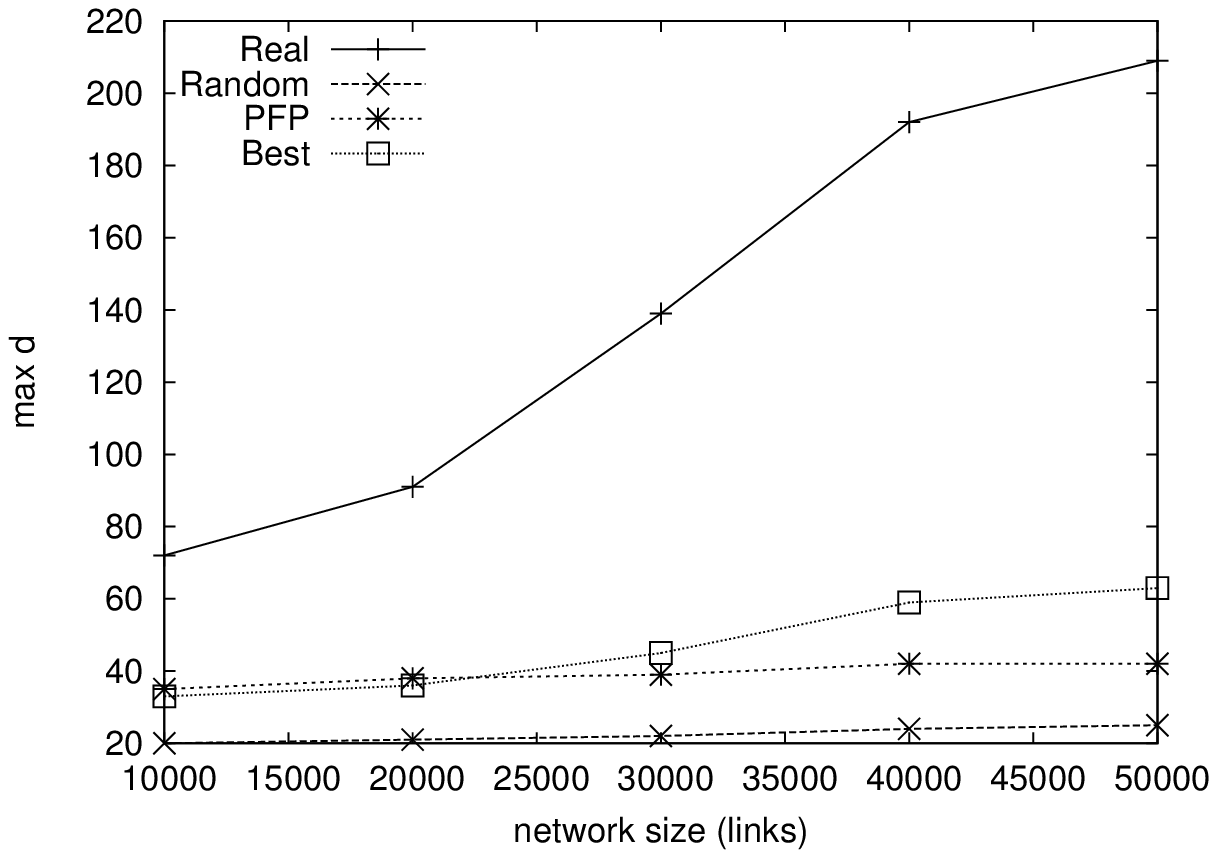} \\
\includegraphics[width=5.5cm]{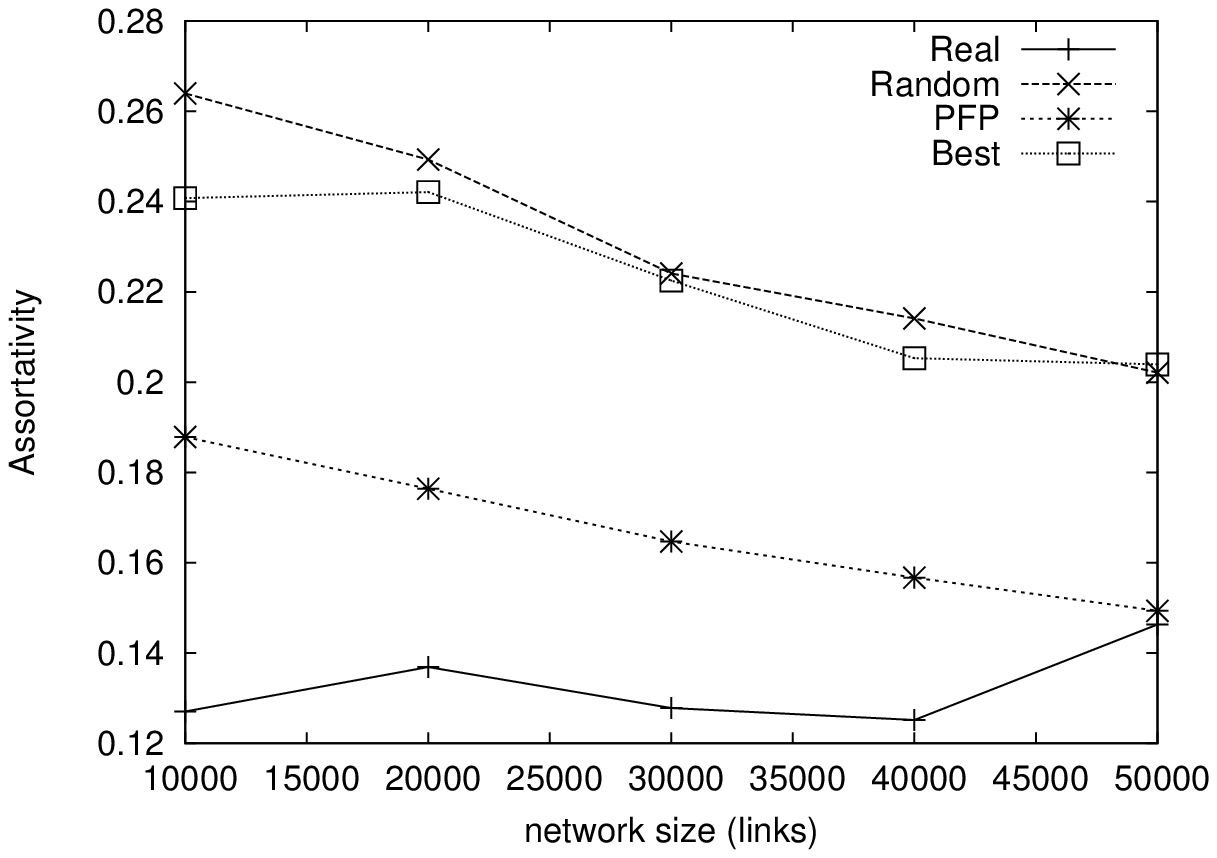}
\includegraphics[width=5.5cm]{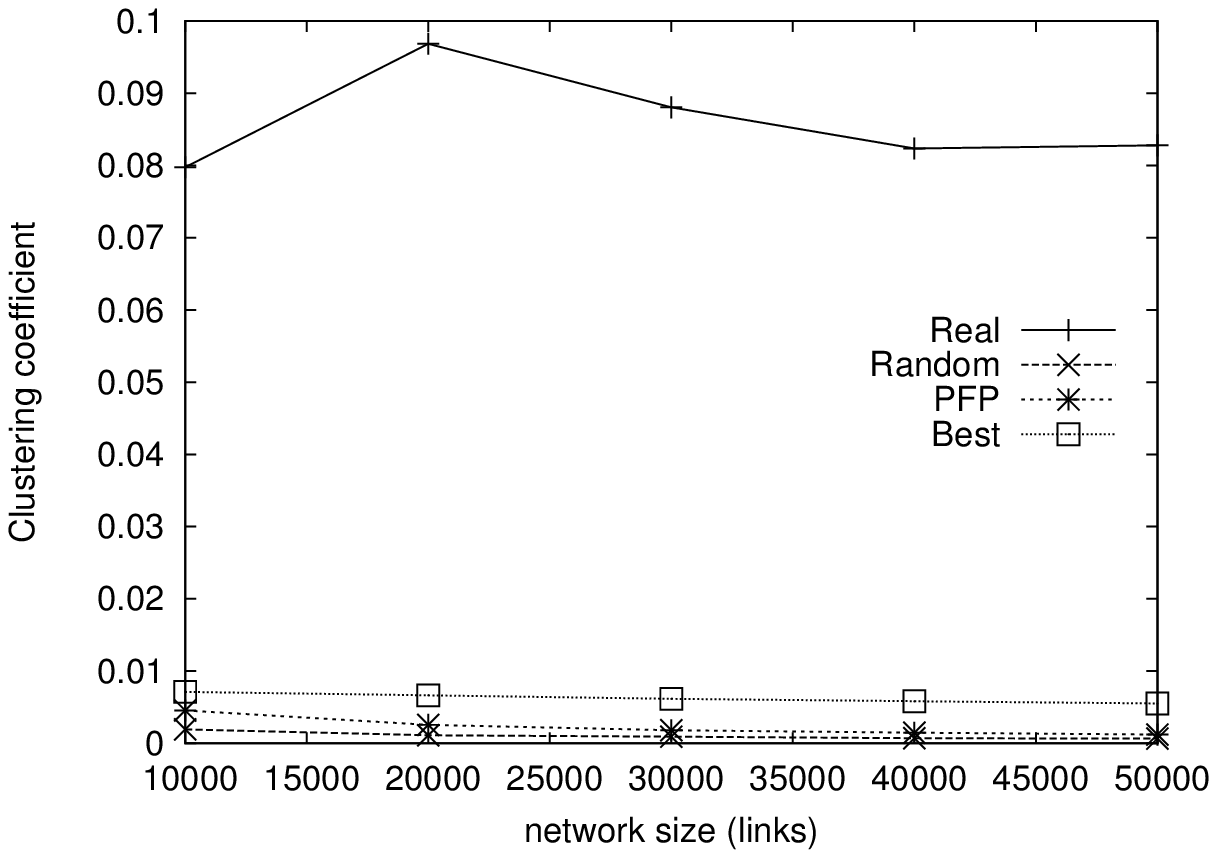}
\includegraphics[width=5.5cm]{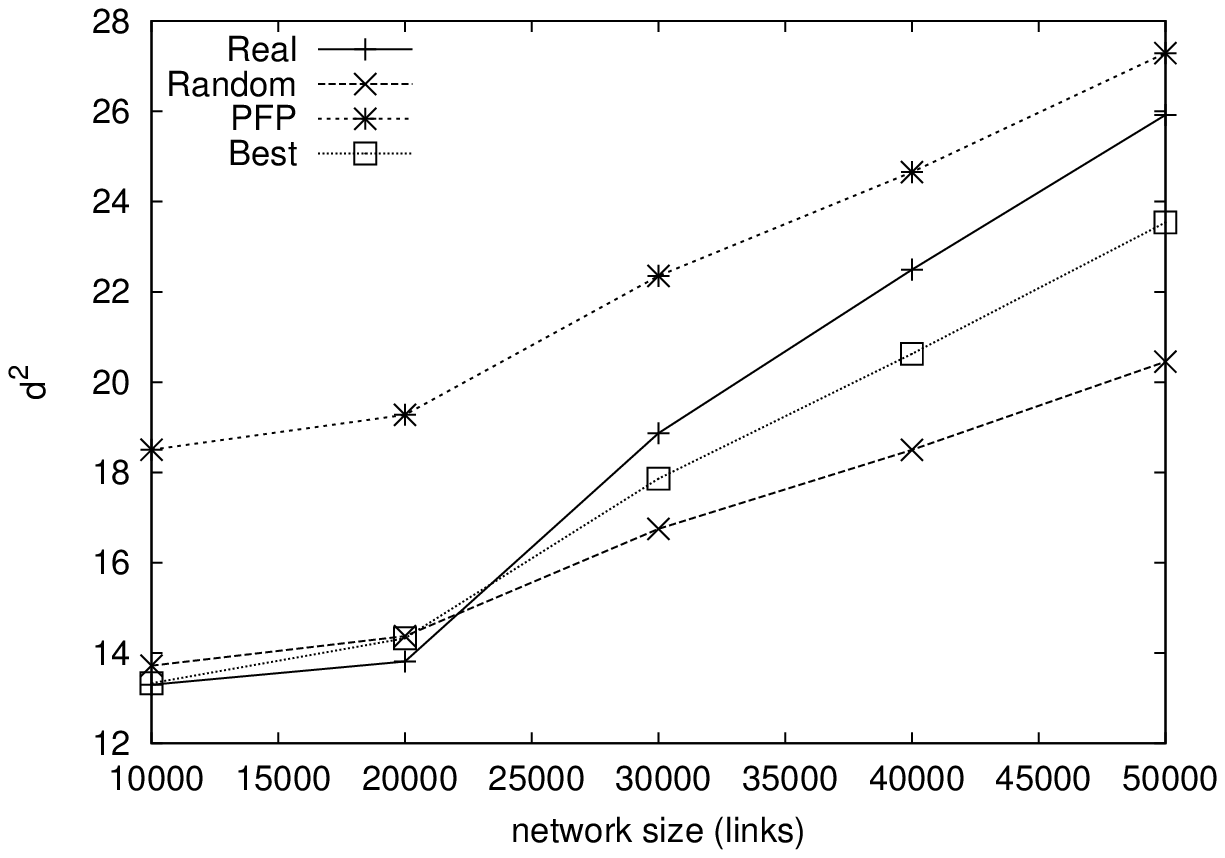}
\caption{Results for gallery network}
\label{fig:gallery}
\end{center}
\end{figure*}

Figure \ref{fig:gallery} shows the results for the gallery network.  From
the $c_0$ values we would expect random to actually be slightly better than
PFP and {\em best\/} to be much better than either.  This order is followed for 
$d_1$, $d_2$ and $\overline{d^2}$ and seems to be for $\gamma$ although all
models are incorrect here.  For assortativity $r$ PFP is unexpectedly the
best model and for $\max d$ it is better than random.  In all cases apart
from $r$ the {\em best\/} model is closest to the real data.  Again the $c_0$ statistic
seems to be a good reflection of the closeness of network statistics, particularly
``first order" statistics.

\subsection{Fitting the Flickr data set}
\label{sec:flickr}

The Flickr\footnote{\url{http://flickr.com/}} website allows users
to associate  themselves with other users  by naming them as {\em Contacts\/}. 
In \cite{Mislove08} the authors describe how they collected data for 
the graph made by users as they connect to other users.  The first 100,000 
links of this network are analysed here.  The graph is generated by a
web-crawling spider so the order of arrival of edges is the order
in which the spider moves between the users rather than the order
in which the users made the connections. Thus, the evolution dynamics 
of this network will be determined in part by the spidering code.

The best pure PFP model for the Flickr data was with $\delta= 0.015$
and this had $c_0 = 28.29$.  The {\em best\/} model had as the model for
new node connections simply $0.99\theta_R(1)+0.01\theta_d$ and 
for connections between existing nodes the best model was 
$0.52\theta_p(-0.22)+0.48 \theta_R(1)$.  This model had the very high
per choice likelihood ratio of 
$c_0 = 430.5$ -- this is because the new node model is almost entirely 
deterministic, new nodes follow by browsing from old nodes.  
It is because the network was from
a browsing pattern that gave the high proportion of $\theta_R(1)$ 
``recent" especially in the new node model.

\begin{figure*}[ht!]
\begin{center}
\includegraphics[width=5.5cm]{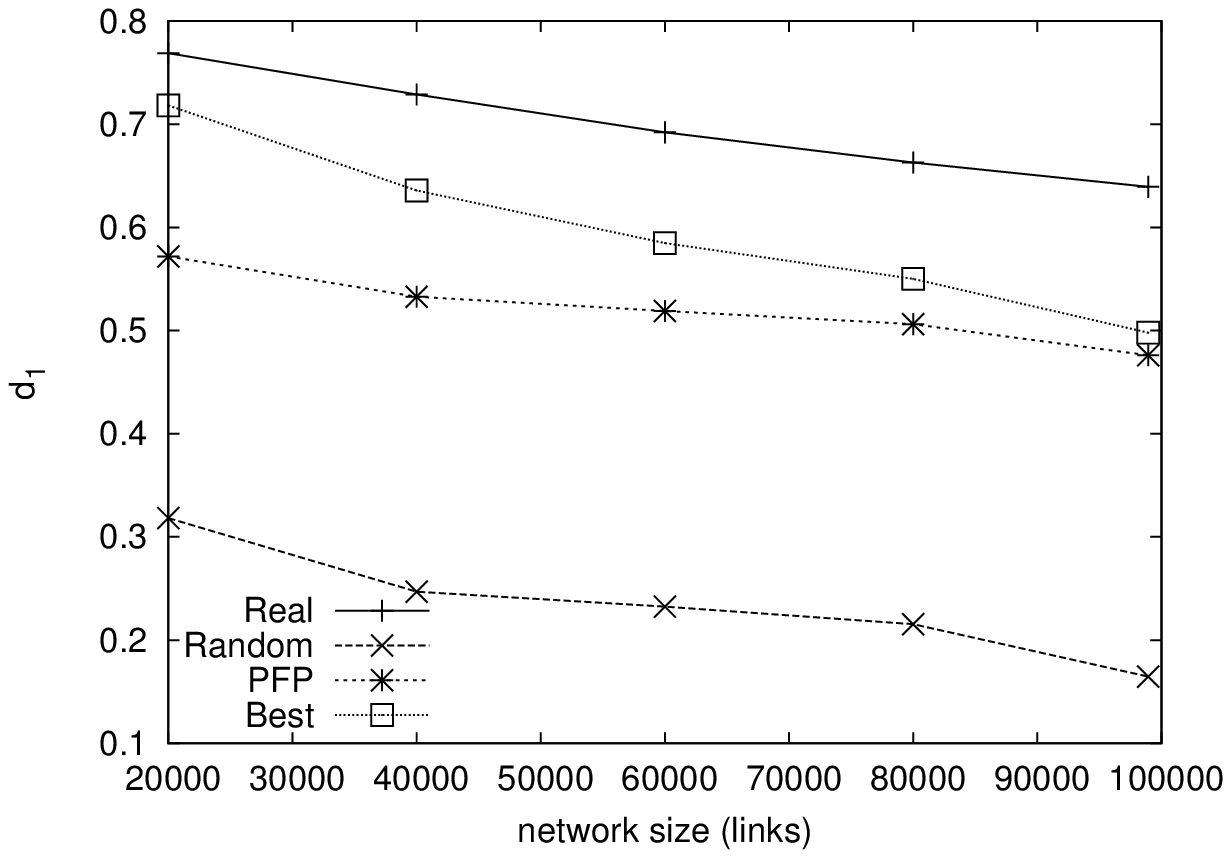}
\includegraphics[width=5.5cm]{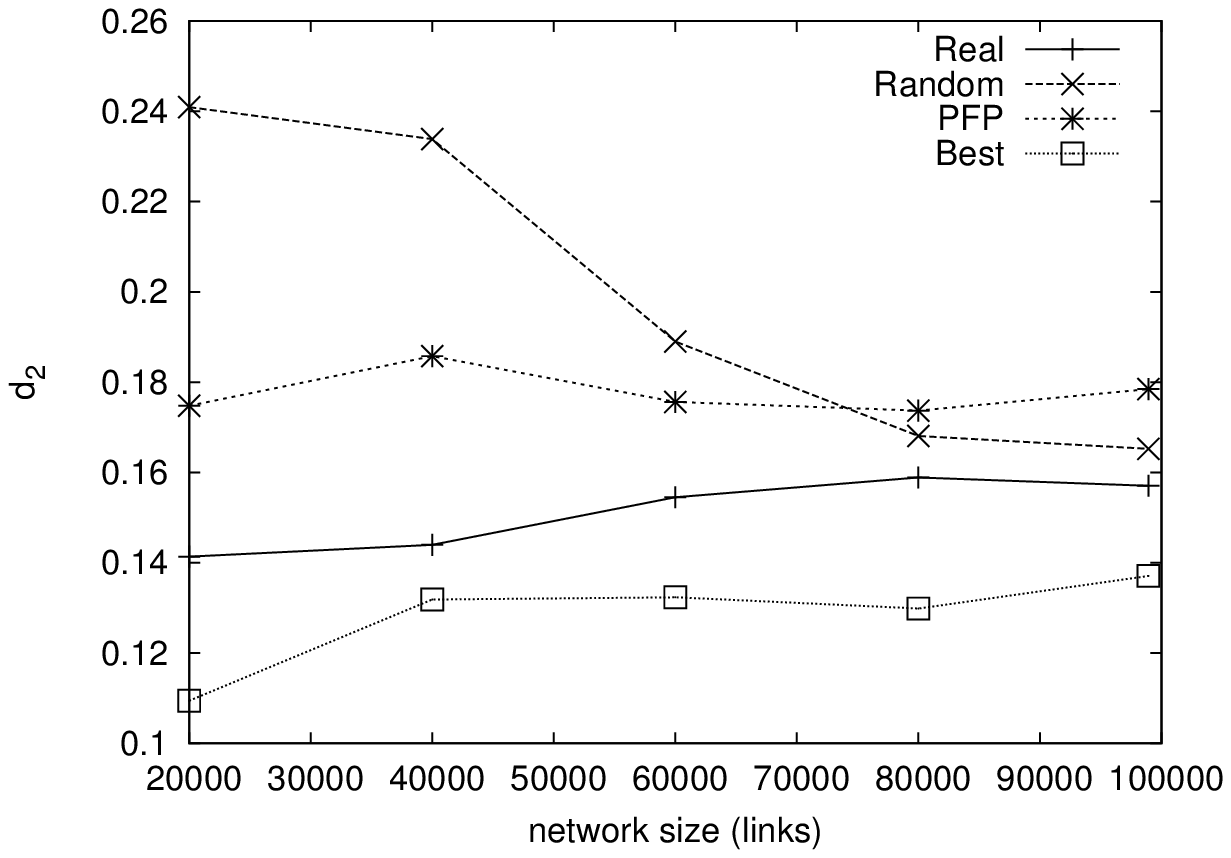}
\includegraphics[width=5.5cm]{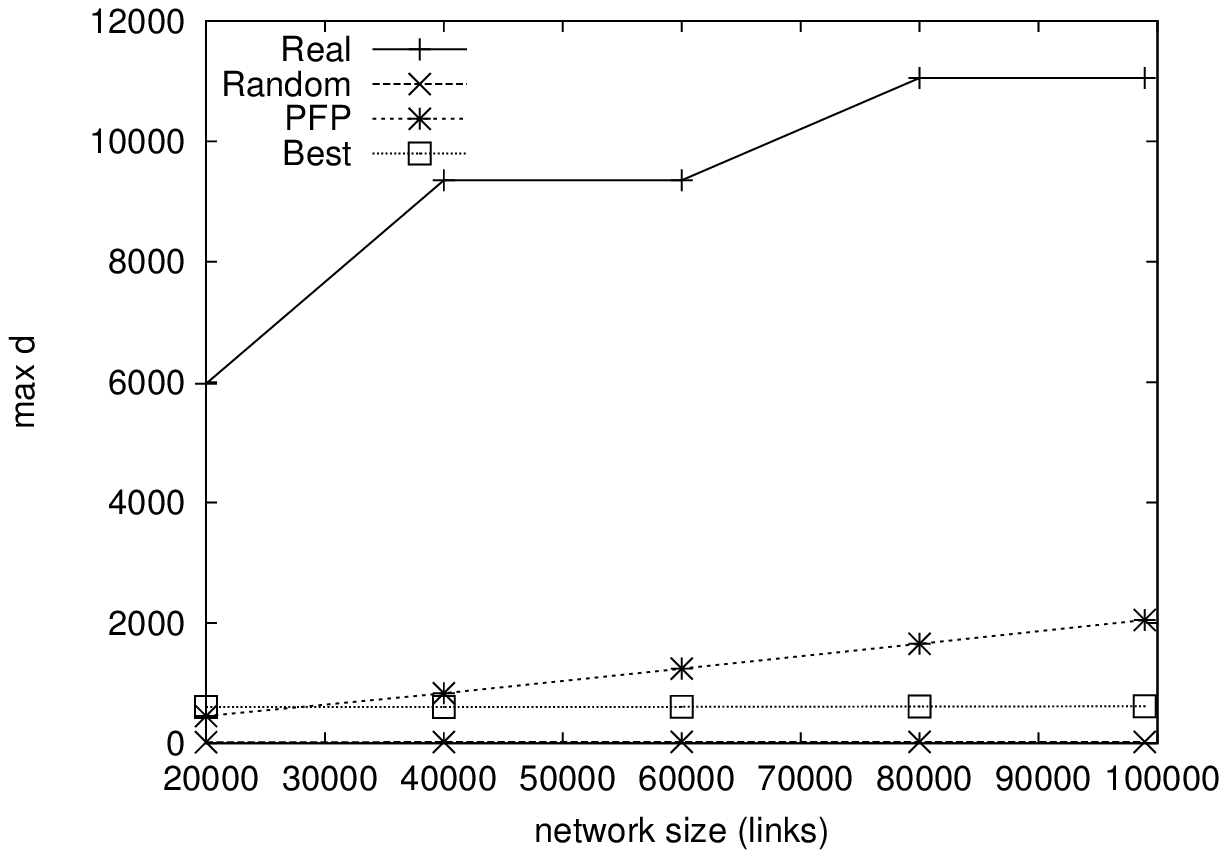} \\
\includegraphics[width=5.5cm]{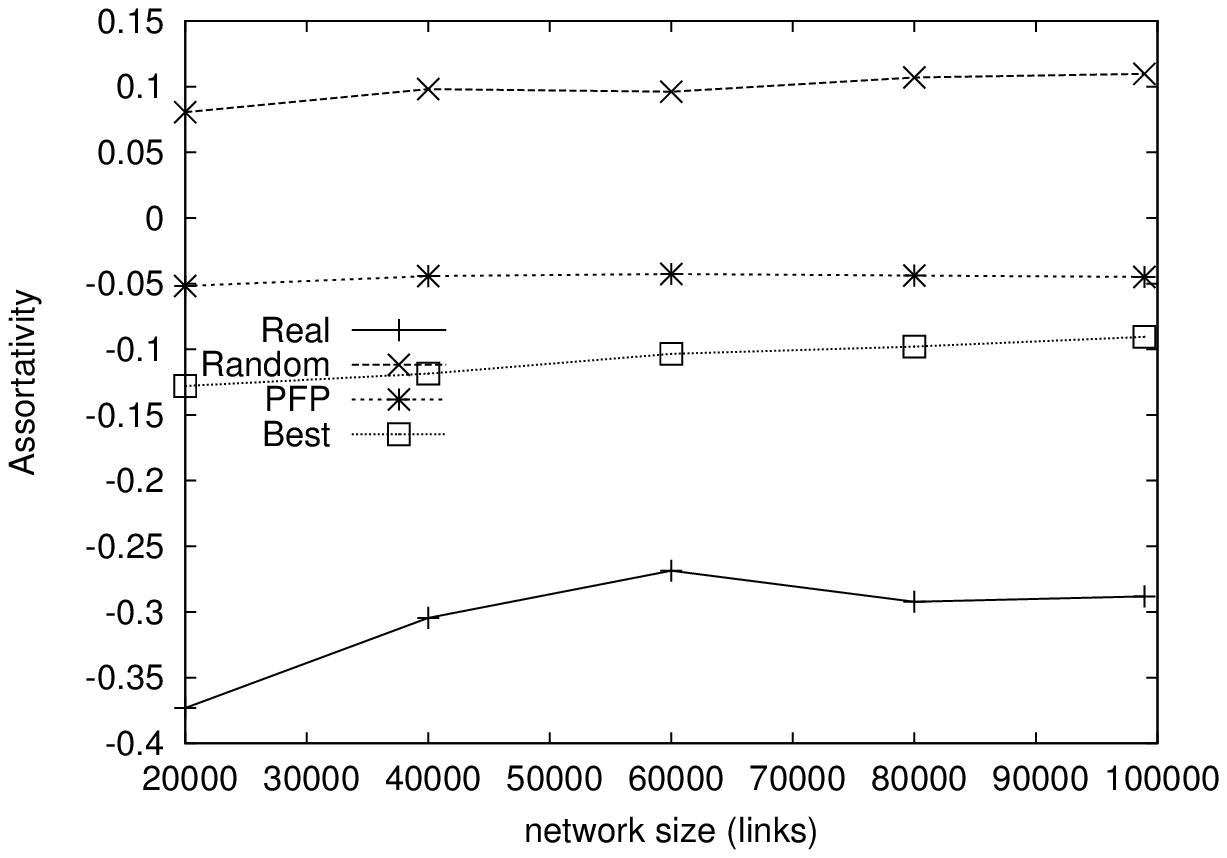}
\includegraphics[width=5.5cm]{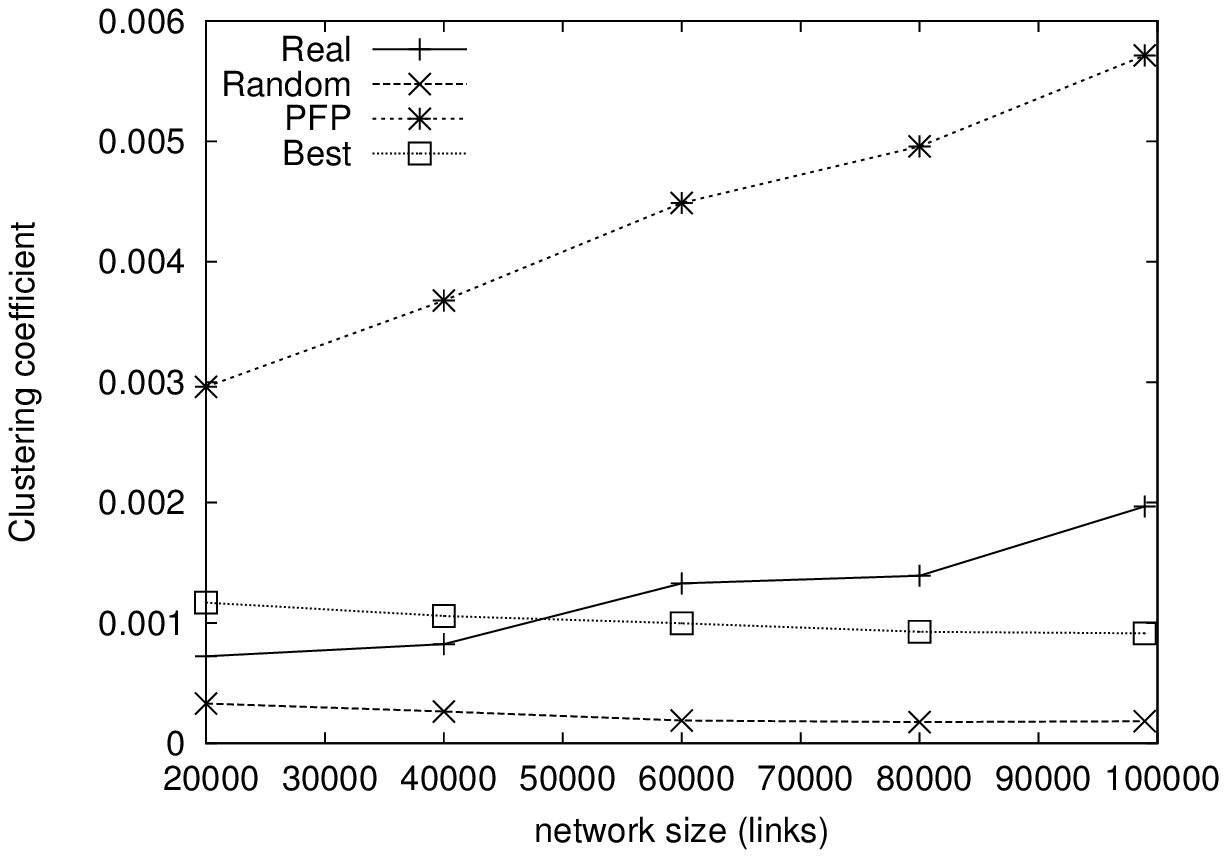}
\includegraphics[width=5.5cm]{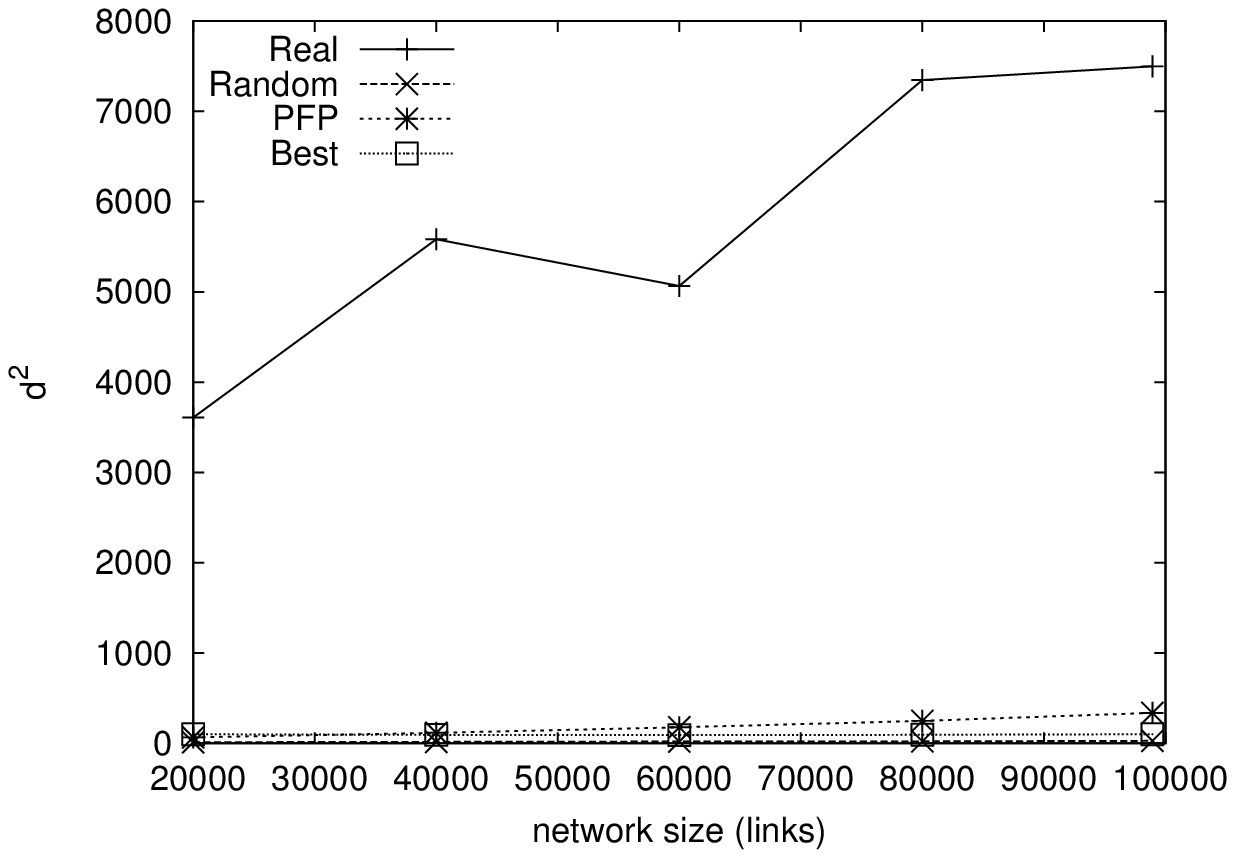}
\caption{Results for Flickr network}
\label{fig:flickr}
\end{center}
\end{figure*}

Figure \ref{fig:flickr} shows the results for the Flickr network.
From the $c_0$ values we would expect the {\em best\/} model to be much 
better than the PFP model which is in turn much better than the
random model.  In fact this is not reflected as strongly in
the statistics as in the previous modelling.  For $d_1$,
$d_2$ and $r$ the statistics are as expected and {\em best\/} is
relatively close.  For $\gamma$, PFP is worse not better
than random.  For $\max d$ and $\overline{d^2}$ no models
are good and the order is not that predicted --
PFP is slightly
better than {\em best\/}.  This may be due to the presence of a single
extremely high degree node (degree 11,053 when the network
has only 46,557 nodes) more than ten percent of the links in
the network are to this single node.

\subsection{Discussion of model fitting}

In general the FETA model assessment performed extremely well
in these tests.  The models were fitted solely with regard
to the likelihood value, without measuring network statistics
in advance.  In all cases, we
believe an impartial observer would rank the models in the same
order as the $c_0$ values.   FETA
was much faster than growing and testing many models.  
A GLM (generalised linear model) procedure
as described in \cite{clegg09} allows optimisation of linear parameters and
dozens of potential sub model combinations can be tested in 
the space of an hour or so.  Growing artificial networks and
testing network statistics can take longer than this to assess a 
single model.  The submodels used focused on first degree node
properties (mainly degree) and this may explain why $\gamma$
and $r$ were not always well fitted.

Some common observations can be made about the models fitted.
PFP and ``recent" were the most commonly used model components.
As expected, PFP models had a negative $\delta$ (sublinear growth)
when the node might be subject to overloading (an author can
only author so many papers a Flickr user can only have time
for a certain number of friends) but positive in systems
where no such overloading was likely (an AS will become more
efficient at adding connections as more people add them).  
The ``spidering" nature of the Flickr data produced an unusual
model for new node connections which were almost always
connected from the most recently connected node, this makes
sense in a ``crawling" environment.  (The likelihood of
$\theta_R(1)$ on its own was zero since at least once this
was not the behaviour observed).  The two AS data sets ended
up with quite similar models which is extremely encouraging
as the fitting was done independently.

\section{Conclusions}
\label{sec:conclusions}
The FETA framework demonstrated in this paper is an excellent way to
test hypothesised models of network evolution if the data set allows
this (evolutionary data must be available).  In the tests here the
model likelihood $c_0$ was an excellent predictor of how close
network statistics would be to those same statistics measured
on the target network.  In addition the statistics usually behaved
in the same way as the network evolved.  The framework has proved
a useful tool for the investigation of five real target networks.

The model components here did a reasonable but far from perfect
job of replicating the real model statistics.  However, the aim
of the paper was to show that the framework could assess models
not to design perfect models.  In this case the models most
common failure was failure to replicate clustering coefficient
and assortativity.  This is perhaps inevitable as the models
were built from components which relied on first order statistics.
Altering the inner model to include second order statistics or
altering the outer model (for example to allow addition of cliques)
could improve this behaviour.

Overall though, the FETA framework is an advance in assessment
of network topology models.  It accounts for the evolution of the
network rather than trying to match a static snapshot.  It provides
a single statistically rigorous likelihood for a model rather than
relying on trying to match a large number of possibly correlated
statistics.  It is computationally cheaper than growing an artificial
test network and measuring statistics to compare with the target network.

Much remains to be done with FETA to improve it.  The outer model needs
attention next and it seems that a similar likelihood procedure would
prove successful here.  Many different sub models can be tried, in particular
focussing on second and higher order statistics seems important.  The
authors welcome collaboration and all software and data used here can be found
at \url{http://www.richardclegg.org/software/FETA}.

\bibliographystyle{abbrv}
\bibliography{nsrl_simplex} 
\balancecolumns 
\end{document}